\documentclass{emulateapj}
\usepackage{graphics,graphicx,epsfig}
\usepackage{amsmath,amssymb,amsfonts} 
\usepackage{subfigure, multirow}
\usepackage{epstopdf}
\newcommand{\lx}{$L_{\rm X}$}
\newcommand{\lir}{$L_{\rm IR}$}
\newcommand{\luv}{$L_{\rm UV}$}
\newcommand{\ergs}{erg s$^{-1}$~}
\newcommand{\ergss}{erg s$^{-1}$}
\newcommand{\ergscm}{erg s$^{-1}$ cm$^{-2}$~}
\newcommand{\lsun}{$L_{\odot}$}
\newcommand{\msun}{$M_{\odot}$~}
\newcommand{\msunyr}{$M_{\odot}$ yr$^{-1}$~}

\newcommand{\eg}{\textit{e.g.,}~}
\newcommand{\ie}{\textit{i.e.,}~}
\newcommand{\GALEX}{\textit{GALEX} }
\newcommand{\Galex}{\textit{GALEX}}

\newcommand{\HST}{\textit{HST} }

\newcommand{\Chandra}{\textit{Chandra}}

\defcitealias{L10}{L10}
\defcitealias{J11}{J11}

\begin{document}

\title{Evidence for Elevated X-ray Emission in Local Lyman Break Galaxy Analogs}
\author{
Antara R. Basu-Zych\altaffilmark{1},
Bret D. Lehmer\altaffilmark{1,2}, 
Ann E. Hornschemeier\altaffilmark{1, 2},
Thiago S. Gon\c{c}alves\altaffilmark{3}, 
Tassos Fragos\altaffilmark{4}, 
Tim Heckman\altaffilmark{2}, 
Roderik A. Overzier\altaffilmark{5}, 
Andrew F. Ptak\altaffilmark{1, 2}, 
David Schiminovich\altaffilmark{6}
}
\altaffiltext{1}{NASA Goddard Space Flight Center, Code 662, Greenbelt, MD 20771; antara.r.basu-zych@nasa.gov}
\altaffiltext{2}{Department of Physics and Astronomy, The Johns Hopkins University, 3400 North Charles Street, Baltimore, MD 21218}
\altaffiltext{3}{Observat\'{o}rio do Valongo, Universidade Federal do Rio de Janeiro, Ladeira Pedro Antonio 43, Sa\'{u}de, Rio de Janeiro - RJ, CEP 22240-060}
\altaffiltext{4}{Harvard-Smithsonian Center for Astrophysics, 60 Garden Street, Cambridge, MA 02138, USA}
\altaffiltext{5}{Department of Astronomy, University of Texas at Austin, 1 University Station C1400, Austin, TX 78712, USA}
\altaffiltext{6}{Department of Astronomy, Columbia University, 550 West 120th Street, New York, NY 10027}

\begin{abstract}
Our knowledge of how X-ray emission scales with star formation at the earliest times in the universe relies on studies of very distant Lyman Break Galaxies (LBGs).  In this paper, we study the relationship between the 2--10~keV X-ray luminosity (\lx), assumed to originate from X-ray binaries (XRBs), and star formation rate (SFR) in UV-selected $z<0.1$ Lyman break analogs (LBAs). We present \Chandra~observations for four new \Galex-selected LBAs. Including previously studied LBAs, Haro~11 and VV~114, we find that LBAs demonstrate \lx/SFR ratios that are elevated by $\sim 1.5 \sigma$ compared to local galaxies, similar to the ratios found for stacked LBGs in the early Universe ($z>2$). Unlike some of the composite LBAs studied previously, we show that these LBAs are unlikely to harbor AGN, based on their optical and X-ray spectra and the spatial distribution of the X-rays in three spatially extended cases. Instead, we expect that high-mass X-ray binaries (HMXBs) dominate the X-ray emission in these galaxies, based on their high specific SFRs (sSFRs~$\equiv$~SFR/$M_\star \geq 10^{-9}$ yr$^{-1}$), which suggest the prevalence of young stellar populations. Since both UV-selected populations (LBGs and LBAs) have lower dust attenuations and metallicities compared to similar samples of more typical local galaxies, we investigate the effects of dust extinction and metallicity on the \lx/SFR for the broader population of galaxies with high sSFRs ($>10^{-10}$ yr$^{-1}$). The estimated dust extinctions (corresponding to column densities of $N_{\rm H}<10^{22}$ cm$^{-2}$) are expected to have insignificant effects on observed \lx/SFR ratio for the majority of galaxy samples. We find that the observed relationship between \lx/SFR and metallicity appears consistent with theoretical expectations from X-ray binary population synthesis models. Therefore, we conclude that lower metallicities, related to more luminous HMXBs such as ultraluminous X-ray sources (ULXs), drive the elevated \lx/SFR observed in our sample of $z<0.1$ LBAs. The relatively metal-poor, active mode of star formation in LBAs and distant $z>2$ LBGs may yield higher total HMXB luminosity than found in typical galaxies in the local Universe.  
\end{abstract}

\section{Introduction}\label{sec:intro}
X-ray binaries (XRBs) dominate the 2--10~keV emission in normal galaxies that do not harbor powerful active galactic nuclei (AGN). In normal star-forming galaxies, high mass X-ray binaries (HMXBs), associated with young ($<100$~Myr) stellar populations, drive the observed correlation between X-ray luminosity and star formation rate  \citep[SFR; \eg][]{Ranalli03,GrimmSFR,BauerXII,HornSDSS, Lehmer08, Mineo12}. 

Recently, \cite{me-lbgstack} found evidence for a mild increase in the mean 2--10~keV luminosity per SFR (\lx/SFR) with redshift for star-forming galaxy populations spanning the majority of cosmic history ($z=0$--5). This result is consistent with X-ray binary population synthesis models, which predict that an increase in \lx/SFR with redshift is expected in response to the declining metallicities of galaxies \citep{Fragos12}. 
   
Currently, the hard X-ray emission in high redshift ($z>2$) galaxies, such as Lyman break galaxies (LBGs), can only be studied in an {\it average} sense by ``stacking" the X-ray counts from large numbers of objects within the deepest \Chandra~surveys \citep[]{BrandtLyBreak,Nandra02,Seibert2002,Lehmer05,Cowie11,Zheng2012, me-lbgstack}. The few LBGs detected individually in the X-ray band at $z\approx3$ are dominated by AGN rather than star-formation related activity \citep{BrandtLyBreak,Nandra02,Lehmer05, Laird06}.~By contrast, LBGs that are not powered by AGN are extremely X-ray faint sources. Observing these sources on an individual basis would require \Chandra~exposures of $\approx8$--20 years, well beyond practical observational limits.

Local ( $D \leq$ 200 Mpc; $z\leq$0.02) starbursts have served as a training set for studying the detailed properties likely to be present in distant LBGs. However, local galaxies differ greatly in the {\it mode} of star formation from LBGs. For instance, LBGs suffer only modest dust extinction in the ultraviolet \citep[UV; e.g., ][]{Webb2003}, whereas local starbursts with comparable SFRs are very dusty with only a few percent of the UV emission escaping \citep[e.g., ][Adelberger \& Steidel 2000]{Heckman1998}. Furthermore, LBGs in the early Universe ($z>2$) formed their stars from more pristine lower metallicity gas than local galaxies \citep{erb06}, potentially producing more luminous HMXBs per SFR \citep{Mapelli09, ZamRob09, Fragos13}.

While the star formation conditions present in LBGs (\ie high SFR, dust- and metal-poor) were commonplace at high redshift, they appear relatively rare in the local Universe. \GALEX all-sky surveys permit the discovery of UV-selected galaxies that are in this currently-rare mode. These local ($z<0.3$) Lyman break analogs (LBAs) are selected to have far-UV properties consistent with $z>2$ LBGs: \eg high SFRs ($\sim$ 10--30 \msunyr), little dust attenuation ($\log L_{\rm IR}$/$L_{\rm UV}\lesssim1$), and relatively low gas-phase metallicities (12$+\log$[O/H]~$< 8.5$). Based on their similarities with LBGs and near proximities, LBAs provide the ideal opportunity to study X-ray emission in {\em individual} LBG-like galaxies. A sample of six $z\sim0.2$ LBAs, targeted for having optical emission-line properties that are intermediate between starbursts and obscured (Type 2) AGN, have been studied by \citet{J11} using XMM and their results suggest the likely presence of Type 2 AGN. However, with the exception of these ``composite'' LBAs and two other individual cases \citep[VV~114 and Haro~11;][]{Grimes06, Grimes07}, X-ray studies of galaxies have not targeted the LBA population comprehensively.
 
%\begin{deluxetable*}{lllcccccc}[b]
\begin{deluxetable*}{lllccccc}[b]
\tabletypesize{\scriptsize}
\setlength{\tabcolsep}{0.0in}
%\tablecolumns{9} 
\tablecolumns{8}
\tablewidth{6.7in}
\tablecaption{Lyman Break Analogs: Summary of \Chandra~ X-ray Observations} 
\tablehead{
  & & & &  \multicolumn{2}{c}{Net Counts}  &  & \\ 
\cline{5-6} 
   &   & \colhead{Exposure} &  \colhead{Aperture} &  &   & \colhead{Flux (2--10~keV)\tablenotemark{a}} & \\
   \colhead{ID} &   \colhead{Obsid \#}  &  \colhead{(ks)} &  \colhead{($\arcsec$)}  & \colhead{0.5--2~keV}& \colhead{2--10~keV} &\colhead{(10$^{-15}$\ergscm)} & \colhead{$\phi^{2-8}/\phi^{0.5-2}$} \\} 
\startdata
J002101.0$+$005248.1   &      13014   &         19.4   &  5  &   26.7$^{+6.4}_{-5.3}$   &    6.8$^{+4.0}_{-2.8}$   &      8.0$^{+4.6}_{-3.2}$   & 0.3$^{+0.2}_{-0.1}$  \\ 
J080619.5$+$194927.3   &     13015   &         20.0   & 5   &    23.8$^{+6.1}_{-5.0}$   &     9.7$^{+4.4}_{-3.3}$    &      10.9$^{+5.0}_{-3.7}$  & 0.4$^{+0.3}_{-0.2}$  \\ 
J082355.0$+$280621.8   &     13012   &      9.0   &   10$\times$20\tablenotemark{*}  &   38.0$^{+7.8}_{-6.8}$   &      14.6$^{+5.4}_{-4.3}$   &  36.7$^{+1.4}_{-1.1}$   & 0.4$^{+0.3}_{-0.2}$ \\
J225140.3$+$132713.4   &    13013   &           19.8   &    5 &   43.3$^{+7.7}_{-6.6}$    & 9.4$^{+4.4}_{-3.3}$   &       10.7$^{+5.0}_{-3.7}$   &  0.2$^{+0.2}_{-0.1}$ \\  
\tablenotetext{a}{2--10~keV  flux assuming simple power law spectrum with $N_{\rm H}=3\times10^{20}$ cm$^{-2}$ (average Galactic extinction for the sample, with variations in this value between different galaxies contributing to $\sim 3$\% uncertainty) and $\Gamma$=1.7. }
\tablenotetext{*}{We used a slightly tilted (position angle of $-7.6^\circ$) elliptical aperture, based on the optical structure, to account for any extended emission in this galaxy. }
\enddata
\label{tab:obs} 
\end{deluxetable*}
\begin{figure*}
\begin{center}
  \includegraphics[width=6.5in]{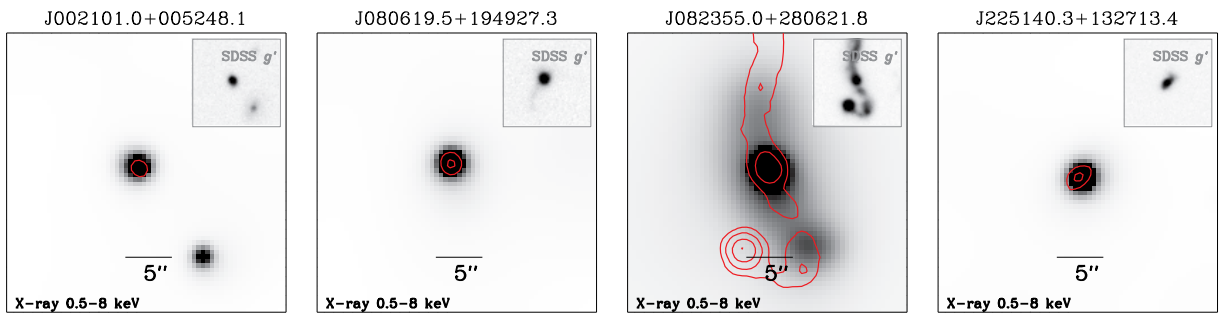}
    \end{center}
  \caption{Shown are full (0.5--8~keV) band X-ray images for the LBA sample with new \Chandra~ observations, overlaid with optical contours corresponding to the SDSS {\it g'} image (shown in the inset at top right of each image, with the same physical scale). All images appear North up and East to left and span 30\arcsec x 30\arcsec. X-ray images have been smoothed using the CIAO {\ttfamily CSMOOTH} package. LBAs are compact sources with half-light radii smaller than 1-2\arcsec. The optical image of J082355.0$+$280621.8 shows complex morphology of interacting galaxies (we show higher resolution \HST archival data for this object in right panel of Figure \ref{fig:LBA3}); the X-ray morphology follows the optical morphology. The bright optical source in the south-east is a foreground star, based on optical spectra taken with the Apache Point Observatory 3.5m telescope.  }
   \label{fig:images}
   \end{figure*}
In this paper, we remedy our limited knowledge of the X-ray properties of LBAs and present \Chandra~observations of six star-formation dominated LBAs. We define the LBA sample in Section \ref{sec:sample} and compare our sample of six LBAs to other larger samples of star-forming galaxies, described in Section \ref{sec:comp}. We find that LBAs have elevated $L_{\rm X}$ per SFR, compared to these other samples, and offer our physical interpretation for this result in Section \ref{sec:results}. We investigate possible physical causes, including the possible presence of AGN (Section \ref{sec:AGN}), the effects of metallicity (Section \ref{sec:metal}), and dust extinction (Section \ref{sec:NH}). We discuss the implications of our results on the high redshift universe in Section \ref{sec:highz}. Section \ref{sec:conclusion} summarizes our results and outlines future plans. Throughout this paper we assume a \citet{Kroupa01} initial mass function (IMF), converting accordingly when comparing with the literature, and adopt a standard $\Lambda$CDM cosmology ($H_0=70$ km s$^{-1}$ Mpc$^{-1}$, $\Omega_{M}=0.3$, $\Omega_{\Lambda}=0.7$). 
   
\section{Data and Analysis}

\subsection{The LBA sample} \label{sec:sample}
The launch of \Galex~enabled rest-frame UV-selection of relatively nearby ($z<0.3$) galaxies in order to discover potential local LBG analogs. \cite{Heckman05} introduced such a sample of UV luminous galaxies (UVLGs) with far UV (FUV; $\lambda \sim 1550\AA$) luminosities, $L_{\rm FUV} >2 \times 10^{10}$~\lsun. This selection was further refined by adding a FUV surface brightness criterion \citep[$I_{\rm FUV} > 10^9$~\lsun~kpc$^{-2}$;][]{Choopes} in order to select ``super-compact UVLGs'' with similar global properties as LBGs. Since this time, several studies have established these galaxies, henceforth called LBAs, as excellent analogs to LBGs \citep{me-radio, me-ifu, rod, overzier2009,Overzier2010, Overzier2011,  tsg, Heckman11-COS}. LBAs are also notably different from other local star-forming galaxies, the former having lower dust attenuations per SFR \citep{me-radio, Overzier2011} and metallicities per stellar mass \citep{overzier2009} than the latter. LBAs are physically compact (half-light diameters $\approx$1--2 kpc;  see contours and inset image in Figure~\ref{fig:images}), often disturbed systems whose SDSS optical spectra indicate that they are starbursts (see Figure~\ref{fig:BPT}). 

We selected a sample of $z<0.1$ LBAs to be studied by \Chandra~by cross-matching the SDSS DR7 spectroscopic and photometric data with the \GALEX GR3 data to find galaxies with $L_{\rm FUV} > 2\times10^{10}$~\lsun~and $I_{\rm FUV} > 10^{9}$~\lsun~kpc$^{-2}$. In this process, we used the SDSS $u$-band half-light radius ($r_{u, 50}$) to approximate the FUV surface brightness ($I_{\rm FUV}\equiv L_{\rm FUV}/\pi r_{u, 50}^2$). To determine $L_{\rm FUV}$, we $k$-corrected the observed \GALEX magnitudes using version 4 of the {\ttfamily k-correct} \citep{kcorrect} IDL code with the 7-band \Galex$+$SDSS (FUV, NUV, $u', g', r', i', z'$) magnitudes. We targeted four of the closest and most UV-luminous LBAs with \Chandra, discussed in the next section (\ref{sec:xray}). In addition, we include in our LBA sample two other well-studied galaxies that are considered good analogs of LBGs: Haro~11 and VV~114. 

\subsubsection{X-ray measurements of LBAs:}\label{sec:xray}
\Chandra~observed four SDSS$+$\Galex-selected LBAs in Cycle 12, using the Advanced CCD Imaging Spectrometer (ACIS)-S camera in the Very Faint observing mode. We used the \Chandra~ Interactive Analysis of Observations (CIAO) package version 4.4.1 with \Chandra~ Calibration Database version 4.5.3 for our analysis. Table \ref{tab:obs} summarizes the \Chandra~ observations for the LBAs. The exposure times (see column 3) ranged from $\sim$9 to 20 ks. We produced exposure maps following the procedure described in \citet[][Section 2.2]{Lehmer2013}. We typically extracted counts within a 5$\arcsec$ aperture to sufficiently include the extent of the optical emission, except in the case of J082355.0$+$280621.8 where we used a 10$\times 20\arcsec$ (radius) elliptical aperture to account for the larger extent of emission in this galaxy (see third panel in Figure \ref{fig:images}). To determine the background counts, we measured the counts within a 15\arcsec~radial aperture at random locations $\sim25\arcsec$ away from our target. In order to ensure that this random background region did not coincide with another source, we repeated this process (\ie selected a new background region randomly) if any pixels within the background region exceeded 2.5$\times$ the mean of the background distribution. The total counts from this background region were scaled by the ratios of the source to background aperture areas to give the net counts. We determined fluxes using PIMMS, assuming a power law with $\Gamma$=1.7 and Galactic extinction, $N_{\rm H}=3\times 10^{20}$ cm$^{-2}$, which is the average value for these four LBAs, using the COLDEN Galactic neutral hydrogen density calculator. The $N_{\rm H}$ values range from 1--5$\times 10^{20}$ cm$^{-2}$, corresponding to negligible ($\sim3\%$) error on the flux calculations. Based on the SDSS spectroscopically determined redshift, we apply a (1+$z)^{\Gamma-2.0}=(1+z)^{-0.3}$ $k$-correction to the 2--10~keV fluxes to calculate 2--10~keV X-ray luminosities (provided in last column of Table \ref{tab:LBAs}). 

\cite{Grimes06} and \cite{Grimes07} have conducted detailed X-ray studies on VV~114 and Haro~11 and we refer to these papers for the information presented in columns 2--4 and 9 of Table \ref{tab:LBAs}. We also compare our sample with a sample of $z\sim0.2$ LBAs, whose X-ray emission was studied by \citet[][henceforth \citetalias{J11}]{J11} using XMM. 

\subsection{Comparison Samples}\label{sec:comp}
\begin{deluxetable*}{ccccccccc}[t]
\tabletypesize{\scriptsize}
\setlength{\tabcolsep}{0.02in}
\tablecolumns{9} 
\tablewidth{6.4in}
\tablecaption{Lyman Break Analogs: Derived Quantities} 
\tablehead{
  &  \colhead{RA}& \colhead{Dec} &  & \colhead{$\log M_\star$} &\colhead{SFR\tablenotemark{a}} &  & & \colhead{$L_{\rm X}$\tablenotemark{c}} \\
 \colhead{ID} & \colhead{(deg)} &  \colhead{(deg)} &  \colhead{$z$} &  \colhead{(\msun)} &  \colhead{ (\msunyr)}  &  \colhead{IRX} &  \colhead{12+$\log$(O/H)\tablenotemark{b} }& \colhead{($10^{40}$ \ergs)} \\
 \colhead{(1)} & \colhead{(2)} &  \colhead{(3)} &  \colhead{(4)} &  \colhead{(5)} &  \colhead{ (6)}  &  \colhead{(7)} &  \colhead{(8)} & \colhead{(9)}} 
\startdata

J002101.0$+$005248.1   &      5.254   &      0.880   &      0.098   &       9.75   &      23.70   &       0.65   &       8.19   &      19$^{+11}_{-8}$  \\ 
J080619.5$+$194927.3   &    121.581   &     19.824   &      0.070   &       9.52   &      15.16   &       0.96   &       8.15   &      13$^{+6}_{-4}$  \\ 
J082355.0$+$280621.8   &    125.979   &     28.106   &      0.047   &       9.46   &      17.11   &       0.88   &       8.23   &      19$^{+7}_{-6}$  \\
J225140.3$+$132713.4   &    342.918   &     13.454   &      0.062   &       9.39   &       8.65   &       0.53   &       8.15   &       10$^{+4}_{-3}$  \\  
  VV114               &     16.946   &    $-$17.507   &      0.020   &      10.65   &      37.82   &        1.0   &       8.45   &      24\tablenotemark{$\dagger$}   \\ 
  Haro11                &      9.219   &    $-$33.555   &      0.020   &      9.84  &      10.88   &       0.52   &       8.33   &      12\tablenotemark{$\dagger$}  \\ 
\tablenotetext{a}{UV$+$IR SFR}
\tablenotetext{b}{We used gas phase metallicities from the SDSS DR7 MPA/JHU VAC \altaffilmark{2}, which were determined using the technique outlined by \citet{VAC_metal}.}
\tablenotetext{c}{2--10~keV X-ray luminosity}
\tablenotetext{$\dagger$}{Other LBAs studied by \citet{Grimes06} and \citet{Grimes07}. }
\enddata
\label{tab:LBAs} 
\end{deluxetable*}
For comparison with the LBAs, we include other samples of ``normal'' (not containing AGN) local star-forming galaxies, studied using \Chandra. These local star-forming galaxies include 24 nearby spiral and irregular galaxies studied by \citet{Colbert04} and 22 additional star-forming galaxies from \citet[][7 of the original 29 galaxies are also in the \citealt{Colbert04} sample]{Mineo12}. In both studies, the galaxies are sufficiently nearby to spatially resolve the XRB population; these authors extracted the X-ray luminosities from XRBs (specifically, HMXBs in the \citealt{Mineo12} sample) separate from the rest of the X-ray emission in these galaxies. 

In addition, our comparison samples include luminous and ultraluminous infrared galaxies (LIRGs and ULIRGs), galaxies which have IR luminosities exceeding 10$^{11}$~\lsun~and 10$^{12}$~\lsun, respectively. These IR-selected galaxies are dusty, high SFR galaxies, whose X-ray properties have been explored in a number of X-ray studies \citep{L10, Iwasawa11, Symeonidis11}. We include 13 LIRGs/ULIRGs from \citet[][hereafter L10]{L10}, after eliminating four likely AGN from their sample of 17, and 29 LIRGs/ULIRGs from the Great Observatories All-Sky LIRG Survey \citep[GOALS;][]{Iwasawa11}, excluding 15 classified AGN from their total sample of 44. While many of the LIRGs/ULIRGs are spatially resolved with \Chandra, the X-ray luminosities refer to galaxy-wide 2--10~keV emission. 

We use the 2--10~keV X-ray luminosities (\lx) given in the references above. In cases where \lx~was derived using a power law with $\Gamma\neq1.7$ (\eg $\Gamma$=1.8; \citealt{Colbert04} and $\Gamma$=2.0; \citealt{Mineo12}), we converted to $\Gamma=1.7$ for the sake of consistency. We note that \lx~values from both LIRG/ULIRG samples (\citetalias{L10} and \citealt{Iwasawa11}) are calculated from fits to the individual galaxy spectra and include a hot-gas component ($kT\lesssim$0.8~keV) plus power-law ($\Gamma=$1-3). Since these are likely to be more accurate than using an assumed power law, we use their quoted values of \lx.

Therefore, the complete sample of comparison galaxies includes 88 galaxies. 

\subsection{UV and IR properties: SFR and IRX}\label{sec:UVIR}
While UV radiation traces recent star formation, it is also easily absorbed by dust and reradiated as infrared (IR) radiation. Therefore a proper accounting of the total star formation rate (SFR) includes both UV$+$IR, as described in the following relation from \cite{Bell05}:
\begin{equation}
\rm SFR (M_\odot yr^{-1}) = 9.8 \times 10^{-11} ({\textit L}_{IR}+3.3 {\textit L}_{UV}),   \label{eqn:sfr}
\end{equation}
where \lir~ and \luv~ correspond to the total IR (8--1000$\mu$m) and UV ($\nu L_{\nu}$ at 2800\AA) luminosities, respectively, in units of solar luminosity. The factor of 3.3 is the typical correction factor that brings $\nu L_{\nu}$ (at 2800\AA) to the integrated $L_{\rm UV}$. All of the LBAs have been observed by the Wide-Field Infrared Survey Explorer \citep[WISE;][]{WISE} and detected in the 22$\mu$m band. We estimated $L_{\rm IR}$ by converting the monochromatic 22$\mu$m luminosity to the total 8--1000$\mu$m luminosity by using the \cite{Charyelbaz} IR SED template. SFRs have been calculated using the \luv~determined from \GALEX data, the \lir~determined from WISE data, and Equation \ref{eqn:sfr}. The amount of dust attenuation in the UV can be approximated by the IR-excess (IRX), defined as
\begin{equation}
\rm IRX = \log ({\textit L}_{IR}/{\textit L}_{UV}).         \label{eqn:irx}
\end{equation} 
SFRs and IRX values for the LBAs are presented in columns 6 and 7 in Table \ref{tab:LBAs}. We estimate SFRs and IRX in a consistent way for the comparison samples using the available values of \lir~and \luv~from the related papers (see section \ref{sec:comp}; given in \citealt{Howell10} for the GOALS sample of LIRG/ULIRGs, and obtained through private communication for \citealt{Mineo12} sample).

\subsection{Stellar mass and Metallicity}\label{sec:logoh}

We estimated stellar masses using 2MASS $K_s$ magnitudes and following the relations given by Table 7 of \cite{Bell_mass}, which are calculated for different choices of optical/NIR colors. The colors used in the stellar mass measurements are SDSS $u'-z'$, $B-K$ and $B-V$ (where $B$ and $V$ photometric data come from the Third Reference Catalog of Bright Galaxies, RC3; \citealt{rc3}) colors for the LBAs, \citet{Colbert04} star-forming and \citetalias{L10} LIRG samples, respectively. \citet{Howell10} use 2MASS $K_s$ magnitudes and IRAC 3.6$\mu$m to estimate $M_\star$ in the GOALS (U)LIRGs, which agrees well (slightly lower by $\sim 4$\%, see \citetalias{L10}) with our method. \citet{Mineo12} estimate $M_\star$ from 2MASS $K_s$ magnitudes, applying the relation given in \citet{Gilfanov04}, and we adjust these values to our adopted IMF (Kroupa).

We measure the metallicities using the method outlined in \citet{PP04}, using the OIII$\lambda 5007$\AA~and NII$\lambda 6584$\AA~emission line ratios (referred to, hereafter, as the PP04 O3N2 method). \citet{KewleyEllison08} compare different metallicity measurement techniques and find that different estimated gas-phase metallicities can differ systematically by 0.7 dex depending on the method. While calibration from various methods to an absolute metallicity value remains uncertain, relative uncertainties are minimized when using one consistent method and the PP04 method proves to be one of the most robust methods \citep{KewleyEllison08}. Emission-line measurements are available for all six of the LBAs and 38 of the 88 comparison galaxies, which we henceforth refer to as the ``partial comparison sample''. Using published emission line values \citep[28 comparison star-forming galaxies, 10 LIRGs, Haro~11, and VV~114; ][]{Kim95, Wu98, MK06,BO02} and SDSS spectroscopic data (four \Chandra-observed LBAs and one LIRG) we determine gas-phase metallicities.

%Emission line measurements were not available for all the galaxies in our comparison sample, but using published emission line values \citep[28 comparison star-forming galaxies, 10 LIRGs, Haro~11, and VV~114; ][]{Kim95, Wu98, MK06,BO02} and SDSS spectroscopic data (four \Chandra-observed LBAs and one LIRG) we determine gas-phase metallicities for 38 comparison sample galaxies and all six LBAs.

The stellar masses and gas-phase metallicities for the LBAs are provided in columns 5 and 8 (respectively) in Table \ref{tab:LBAs}. 
  
\section{Results}\label{sec:results}

\subsection{Elevated \lx/SFR in UV-selected galaxies}

   \begin{figure}
\begin{center}
  \includegraphics[width=3.5in]{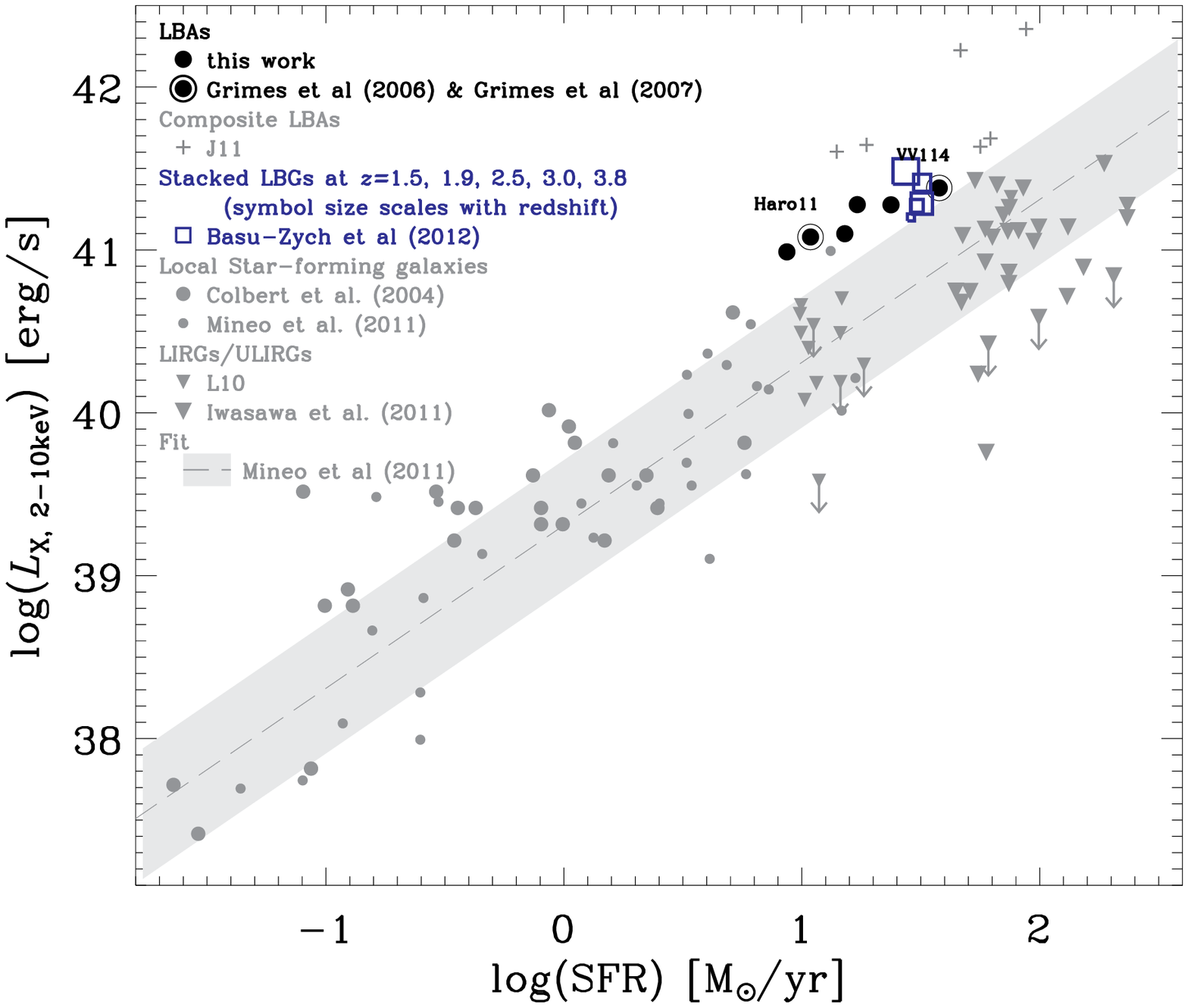}
    \end{center}
  \caption{The LBAs (black points) and stacked LBGs (open squares; \citealt{me-lbgstack}) are significantly more X-ray luminous (in 2--10~keV band) per SFR compared to other galaxy populations: local galaxies (large gray points from \citealt{ColbertXLF2003} and small gray points from \citealt{Mineo12}) and LIRGs/ULIRGs (filled gray triangles from \citealt{Iwasawa11} and \citetalias{L10}). The stacked LBGs (open squares; \citealt{me-lbgstack}) and LBAs, appear similarly high with respect to the 2--10~keV Xray-SFR correlation (gray dotted line shows fit to data from \citealt{L10}, dotted line shows the extrapolated fit based on the low SFR galaxies). We show the J11 sample of composite LBAs as crosses for comparison. }
   \label{fig:lxsfr}
   \end{figure}
   
\begin{figure*}
\begin{center}
  \includegraphics[width=6.5in]{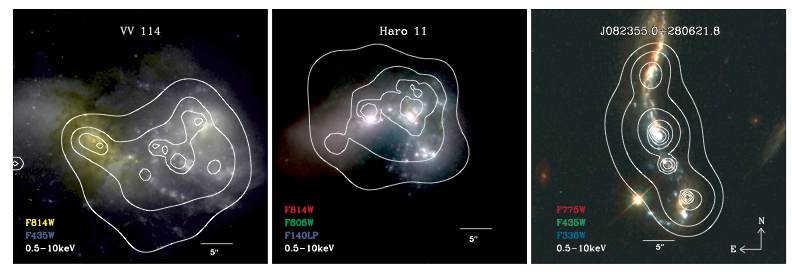}
    \end{center}
  \caption{We present archival \HST images of VV~114 ({\em left}; PI: A. Evans), Haro11 ({\em center}; PI:G. \"{O}stlin) and LBA J082355.0$+$280621.8 ({\em right}; PI:G. \"{O}stlin) with white contours showing the \Chandra~full-band (0.5--10~keV) distribution, smoothed using {\ttfamily CSMOOTH}. We are able to resolve multiple X-ray sources with the \Chandra~data for these sources, and find a higher number of \lx$>10^{40}$ \ergs ULXs than expected, based on the HMXBs in typical star-forming galaxies.}
   \label{fig:LBA3}
   \end{figure*}

We show the relation between the 2--10~keV X-ray luminosity and SFR (\lx~vs. SFR) for the LBAs compared to the full star-forming sample of 88 galaxies in Figure \ref{fig:lxsfr} (see legend for symbol descriptions). On average, the UV-selected samples (LBAs and the stacked $z=1.5-4$ LBGs) appear to have \lx~per SFR elevated by a factor of $\sim 1.5\sigma$ compared to the comparison sample. The likelihood that all six LBAs would randomly scatter this far from the local relation is $\sim 10^{-5}$ (based on $\chi^2$), almost certainly implying there is something fundamentally different about LBAs compared to typical normal galaxies. 

The J11 sample of $z\sim0.2$ LBAs, having optical emission-line signatures between starbursts and obscured AGN (see further discussion in section \ref{sec:BPT}), are shown in Figure \ref{fig:lxsfr} as crosses. Two of the J11 sources have \lx/SFR that are extreme (factors $\sim10$--20 above the relation), likely suggesting that AGN power the X-ray emission in these sources; two other sources appear similarly elevated (by $\sim 1.5\sigma$, or factor of $\sim 4$) as the LBAs in our sample. 

Several studies have shown that 2--10 keV emission in local galaxies correlates well with SFR \citep[\eg][]{Ranalli03, PR2007, L10, Mineo12}. The 2--10~keV emission mainly originates from accreting X-ray binaries (XRBs, including ultraluminous X-ray sources, ULXs).  Although minimized by studying emission in the 2--10~keV band, we note that a number of other sources can also contribute to X-ray emission in normal galaxies: supernovae and their remnants, and hot gas from starburst-driven winds and outflows \citep[see, \eg review by][]{Fabbiano89}. Low-luminosity AGN activity may also contribute to the 2--10~keV X-ray luminosity in normal galaxies. In the next section, we assess the likelihood that the presence of AGN cause the elevated \lx/SFR observed in LBAs.

\subsection{Assessing any possible AGN contribution}\label{sec:AGN}
Since AGN also produce significant X-ray emission, we explore the possibility that hidden AGN may cause \lx~values per SFR that are $\sim1.5\sigma$ higher in the LBAs. Benefitting from the high angular resolution capability of \Chandra, three of the six LBAs in our sample, VV~114, Haro~11, and J082355.0$+$280621.8 appear to have extended X-ray emission (see Figure \ref{fig:LBA3}), and we can distinguish multiple X-ray point sources, all of which are luminous enough to qualify as ULXs (\lx$>10^{40}$\ergs). The nuclear regions, potentially associated with AGN, contribute only $<$15\% %[CHECK THIS] 
to the 2--10~keV emission in each case, immediately indicating that the elevated \lx/SFR for those LBAs cannot be attributed to AGN activity. 

While VV114 may harbor an obscured AGN (in the Eastern component, VV114E), \citet{Grimes07} conclude that star formation dominates the 2--10~keV power in this galaxy. Based on {\em Spitzer}/IRS spectroscopic analysis of the [OIV]$\lambda25.9\mu$m and [NeV]$\lambda14.3\mu$m emission features, \citet{Cormier2012} conclude that Haro~11 does not contain an AGN. To further test for the presence of AGN in the four newly-observed SDSS$+$GALEX-selected LBAs, we use optical spectral line diagnostics and X-ray spectral constraints (see below).

\subsubsection{Optical spectral line diagnostics}\label{sec:BPT}
\begin{figure*}
\begin{center}
  \includegraphics[width=6.in]{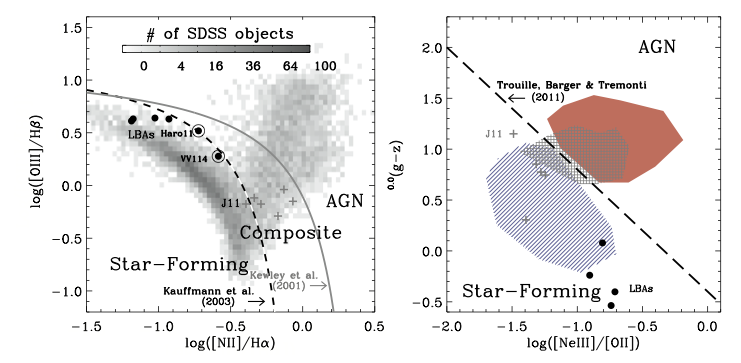}
    \end{center}
  \caption{We use two different AGN vs. star formation diagnotics: BPT diagram ({\em left}) and TBT diagram ({\em right}; \citealt{TBT}). {\em Left:} The full SDSS DR7 galaxy sample is shown (background data) with $z<0.1$ LBAs (black points). These emission-line ratios are determined after subtracting off template galaxy spectra and are very sensitive to even low-level AGN activity. The two lines in the diagram indicate the theoretical upper limit for star formation \citep[gray upper solid line;][]{Kewley01} and the lower line below which AGN are not expected \citep[black dashed line]{Kauff2003AGN}. {\em Right:} The TBT diagnostic separate AGN (above dashed line) from star-forming galaxies. The LBA sample falls well within the ``star-forming'' space for both diagnostics. We have also used velocity widths to screen against AGN. The \citetalias{J11} sample of ``composite LBAs'' is shown as crosses.}
   \label{fig:BPT}
   \end{figure*}

The Baldwin, Phillips, and Terlevich \citep[BPT;][]{BPT} diagram separates galaxies whose optical emission is dominated by AGN versus star formation using the ratios of forbidden lines (\ie [NII]$\lambda 6584$\AA, [OIII]$\lambda 5007$\AA), which are excited mainly by AGN, to nearby Balmer lines. In the left panel of Figure \ref{fig:BPT}, we show the distribution of $>900,000$ sources from the SDSS DR7 catalog (shaded background). The solid gray line is based on a theoretical modeling of the upper limit that star forming galaxies can occupy in this diagram by \citet{Kewley01}. \citet{Kauffmann03AGN} revise the classification criterion to further separate star-forming from composite galaxies. The LBAs in this study (black points) reside in the star-forming region of this diagram. 
 
Using XMM, \citetalias{J11} studied a sub-sample of six $z\sim0.2$ LBAs (shown as crosses in Figures \ref{fig:lxsfr}, \ref{fig:BPT}, and \ref{fig:lxsSFR}) that lie within or near the composite region of the BPT diagram. They also found that these galaxies are systematically offset with respect to the X-ray/SFR relation, having  \lx$= 4$--20$\times 10^{41}$ \ergs (see Figure \ref{fig:lxsfr}). Based on comparing the X-ray luminosities with the far-IR and [OIII]$\lambda$5007\AA~luminosities, J11 conclude that Type 2 AGN are likely present within their sample. One of the composite LBAs most offset in \lx~was also shown to host a compact radio source using VLBI, and interpreted as an AGN having a fraction of the total radio flux of the LBA \citep{Alexandroff}. Other LBA 'composites' were not detected at VLBI resolution, showing that their overall radio emission is dominated by star formation.

Motivated by the study by \cite{HornSDSS} which showed that using the BPT diagram alone may still include AGN (\eg narrow-line Seyfert 1 galaxies), we confirmed that our sample of LBAs had Balmer line widths that were consistent with those of the forbidden lines (within 3$\sigma$ compared to the distribution of the entire SDSS DR7 population).

Another emission line test has been proposed by \cite{TBT}, referred hereafter by the TBT test (shown on right panel of Figure \ref{fig:BPT}), which uses rest-frame $g-z$ color and the ratio of [NeIII]$\lambda3869$\AA~to [OII]$\lambda\lambda3726$\AA,3729\AA~to separate AGN (upper region) from star-forming galaxies (below dashed line). \citet{TBT} compare their method with the BPT diagnostic; the solid red, dashed blue, and cross-hatched gray regions in Figure \ref{fig:BPT} [right] correspond to the BPT-determined AGN, star-forming, and composite classifications, respectively. According to this classification, the LBAs reside in the star-forming region of the TBT diagram. 

The TBT method is good for selecting sources that might be X-ray detected. In their analysis, \citet{TBT} find that X-ray sources with \lx~$>10^{42}$~\ergs appear in the top region, while few luminous X-ray sources fall below the dashed line. Moreover they show that the TBT method is able to classify luminous, less luminous, and soft X-ray detected non-broad line AGN sources accurately, while $\sim 20$\% of X-ray detected AGN are classified as star-forming galaxies. We note that the J11 ``composite LBAs'' (crosses), based on the BPT classification, were detected at \lx~$\sim10^{42}$~\ergss, yet fall in the star-forming region of the TBT classification. 

Our sample of LBAs lies conservatively within the star-forming region of the TBT diagram, providing further support against AGN. While using optical emission lines to detect AGN is expected to be sensitive to even low-levels of AGN activity, this method may suffer from aperture dilution in some of the higher redshift sources or may miss heavily obscured AGN. 

\subsubsection{X-ray Spectral properties of LBAs} \label{sec:Xrayspec}
We test for the presence of obscured AGN, whose optical spectral signatures may be hidden due to attenuation by dust. Assuming a power law spectrum, the X-ray hardness ratio (defined here as the ratio of emission in the 2--10~keV band to that in the 0.5--2~keV band, H/S) is affected by the X-ray absorption column, $N_{\rm H}$, and the slope of the power law, $\Gamma$. The hardness ratios are listed in Table \ref{tab:obs}, and range between 0.2--0.4, suggesting very soft X-ray spectra ($\Gamma=2.8$--3.3). 

While we do not have enough counts to fit the X-ray spectra for individual LBAs, we can perform a joint fit to the four $\Chandra$-observed LBAs to obtain an average spectral shape for the sample. Even combining the data from four sources does not provide the adequate number of counts to constrain a spectrum with multiple components and therefore we restrict our analysis to fitting a simple power law. For the 0.5--8~keV band, we find that the best fit power law has $\Gamma_{0.5-8}=1.91\pm0.15$, consistent with XRB spectra and inconsistent with heavily-obscured and unobscured low-luminosity AGN with $L/L_{\rm Edd}\lesssim 0.1$ \citep[\eg][]{Shemmer06}. 

Additionally, the X-ray emission in starburst galaxies may contain a significant hot gas component, which would contribute strongly to the soft band, as found in the other LBAs, Haro 11 \citep{Grimes07} and VV 114 \citep{Grimes06}. Although we are not able to add a thermal component to our fit, we find that the best fit power law hardly changes when we fit the 1--8~keV band, $\Gamma_{1-8}=1.89\pm0.21$. 

In their Figure 16, \citet{Xue11} study X-ray sources from the 4~Ms \Chandra~Deep Field-South (CDFS) survey to show that AGN are unlikely to have this combination of low 0.5--8~keV X-ray count rate ($<10^{-3}$ count s$^{-1}$) and soft X-ray spectra ($\Gamma > 1.5$). Therefore, it is unlikely that all six LBAs harbor unobscured low-luminosity AGN. %However, sources having such low X-ray luminosities ($<10^{42}$ \ergs) combined with relatively soft spectra ()  are rarely observed \citep[\eg see large crosses and triangles in Figure 16 in][which shows the ]{Xue11}, indicating it is unlikely that all six LBAs harbor unobscured low-luminosity AGN. 

Based on the separate tests for the presence of AGN discussed in this section, we conclude that it is unlikely that AGN provide significant contributions to the 2--10~keV emission for the LBAs in our sample. %We focus our investigation on how XRB populations in the LBAs drive the elevated \lx/SFR. 

\section{Discussion}
Given the unlikely explanation that AGN are causing the elevated \lx/SFR in the LBAs, we focus on the XRB populations within these galaxies to address the question -- could differences in either the XRB environments (\ie due to effects that elevate \lx/SFR at lower metallicities) or observing conditions (\ie less dust attenuation absorbing the X-ray emission), between the LBAs and typical star-forming galaxies in the local Universe, cause the $\sim 1.5\sigma$ elevated \lx/SFR ratios that are observed in LBAs? First, we discuss how \lx/SFR relates to the XRB population within galaxies. 
\subsection{HMXBs in star-forming galaxies}\label{sec:sSFR}
   \begin{figure}[t]
\begin{center}
  \includegraphics[width=3.1in]{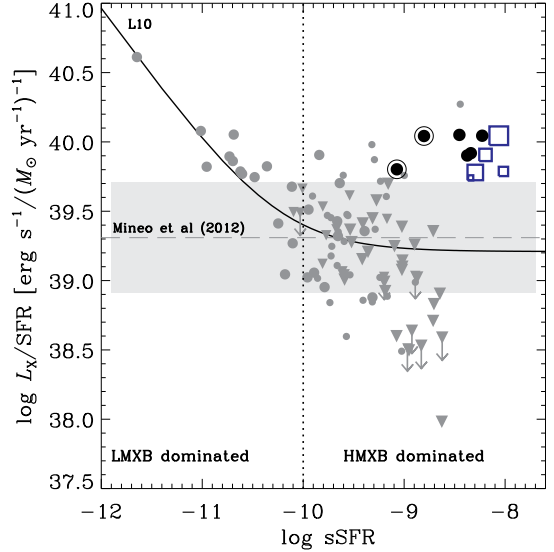}
    \end{center}
  \caption{We compare \lx/SFR versus sSFR for LBAs (black points), LBGs (blue squares, symbol size increasing with sample redshift from $z=1.5$--4) and other star-forming galaxies (symbols are described in Figure \ref{fig:lxsfr}). Specific SFRs (sSFRs) separate galaxies whose X-ray luminosity is dominated by LMXBs (left of the dashed line) from those dominated by HMXBs (with sSFR $>10^{-10}$\msunyr). The solid black line shows the fit from \citetalias{L10}, and the gray shaded region and dashed line show the HMXB-driven relation from \citet{Mineo12}. Similar to LIRGs/ULIRGs, LBAs have high sSFRs, and therefore \lx~in LBAs is expected to be dominated by HMXBs. However, LBAs appear to have elevated \lx/SFR, compared to other high sSFR galaxies (including the LIRGs and ULIRGs), potentially driven by their lower metallicities.}
   \label{fig:lxsSFR}
   \end{figure}

In Figure \ref{fig:lxsSFR}, we show the relationship between \lx/SFR and specific SFR (sSFR$={\rm SFR}/M_\star$) for the LBAs (black points), stacked $z=1.5$--4 LBGs (blue squares) and comparison sample (gray symbols). LBAs, LBGs and LIRGs/ULIRGs (gray triangles) have similarly high sSFRs, but \lx/SFR is $\sim 1.5\sigma$ higher for the LBAs, suggesting that the XRB population in LBAs may be unique, even compared to other galaxies that have high sSFRs (see Section \ref{sec:metal}). 

Within the XRB population, high-mass X-ray binaries (HMXBs) are short-lived, tracing recent star formation activity (on timescales $\sim10^{6-7}$ yrs), while low-mass X-ray binaries (LMXBs) trace older stellar populations (for timescales $>10^{8-9}$ yrs) and the 2--10 keV X-ray luminosity scales with the stellar mass ($M_\star$) of the galaxy. Assuming the following analytic parameterization for local (within a distance of 60 Mpc) normal galaxies:  
\begin{eqnarray}
L_{\rm X} & = & L_{\rm X}({\rm LMXB}) + L_{\rm X}({\rm HMXB}) \label{eqn:lx} \\
&=& \alpha M_\star  + \beta {\rm SFR} \label{eqn:lxmasssfr} 
\end{eqnarray} 
\citetalias{L10} measured constants of $\alpha=(9.05 \pm 0.37)\times 10^{28}$ \ergs~M$_\odot^{-1}$ and $\beta=(1.62\pm0.22) \times 10^{39}$ \ergs~ (\msunyr)$^{-1}$ \citep[see also][]{Colbert04}.  Rearranging Equation \ref{eqn:lxmasssfr} to give \lx/SFR in terms of sSFR (${\rm SFR/M_\star}$) ~\citetalias{L10} show that $\log$ \lx/SFR is proportional to $\log (1/{\rm sSFR} + constant)$, causing two regimes: at low sSFRs ($\leq10^{-10}$ yr$^{-1}$), where both LMXBs and HMXBs contribute significantly to the X-ray luminosity, the relation is inversely linear; at high sSFRs ($\geq10^{-10}$ yr$^{-1}$), where the contribution of LMXBs is negligible and the X-ray luminosity scales with SFR, the relation flattens (see black curve in Figure \ref{fig:lxsSFR}).

 \citet{Mineo12} correct the observed 0.5--2~keV luminosities for contributions from LMXBs and hot gas in their sample of star-forming galaxies (smaller gray points), and fit for the HMXB-driven \lx/SFR relation. After converting their relation to 2--10~keV, using $\Gamma=1.7$, we show the HMXB-driven X-ray/SFR relation by the gray dashed line (shaded region marks the 1-$\sigma$ scatter) in Figures \ref{fig:lxsfr}, \ref{fig:lxsSFR}, and \ref{fig:NH}. Hereafter, we refer to the relation as the local \lx/SFR relation and restrict our analysis to high sSFR, HMXB-dominated galaxies, with sSFRs$>10^{-10}$ yr$^{-1}$ (right of the dotted line in Figure \ref{fig:lxsSFR}), to simplify the discussion by focussing on the relationship between galaxy properties and the HMXB population. 

\subsection{Metallicity dependence}\label{sec:metal}
\begin{figure}[t]
\begin{center}
  \includegraphics[width=3.1in]{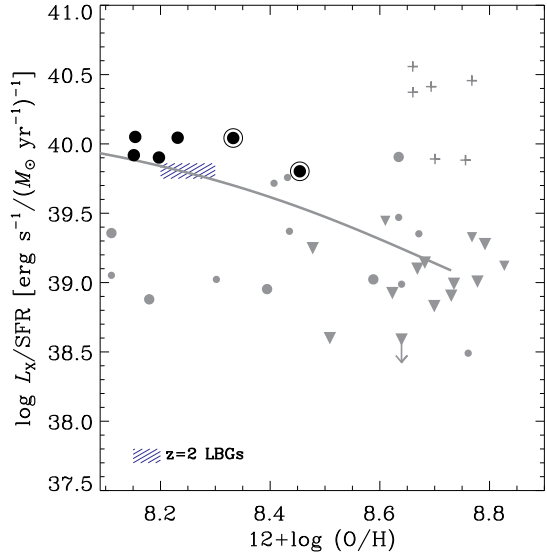}
    \end{center}
  \caption{\lx/SFR versus gas-phase metallicity for galaxies with sSFR$>10^{-10}$ yr$^{-1}$, whose X-ray luminosities are expected to be dominated by HMXBs. The symbols are the same as in Figure \ref{fig:lxsfr}, with the addition of a blue hatched region, corresponding to $z\sim2$ LBGs (see text for details). There is a correlation between \lx/SFR and gas-phase metallicity ($12 + \log [\rm O/H])$, as predicted for HMXBs based on XRB population synthesis models \citep[gray curve;][]{Fragos13}.  }
   \label{fig:metal}
   \end{figure}

\begin{figure*}
\begin{center}
  \includegraphics[width=3.in]{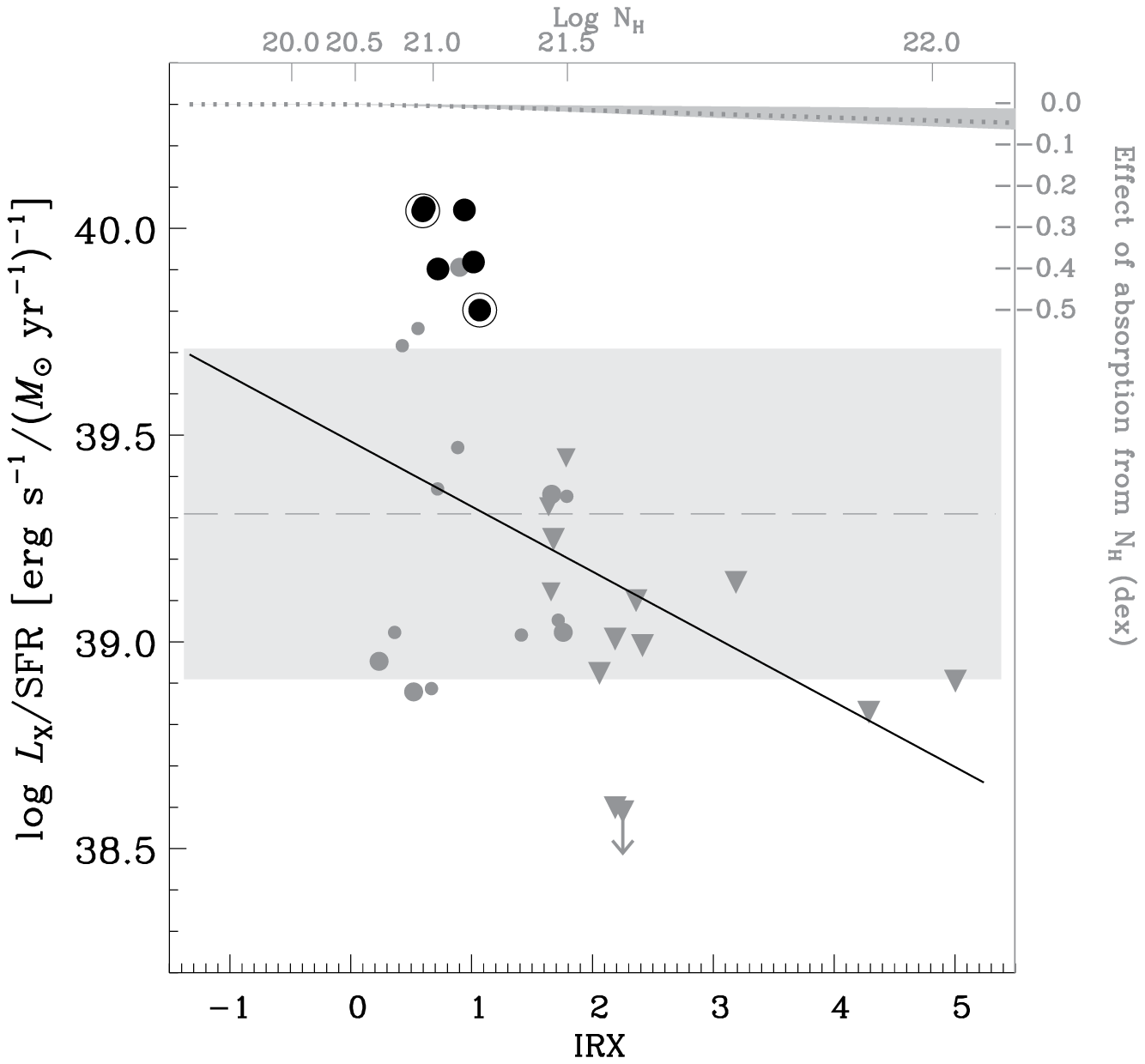}
  \includegraphics[width=3.in]{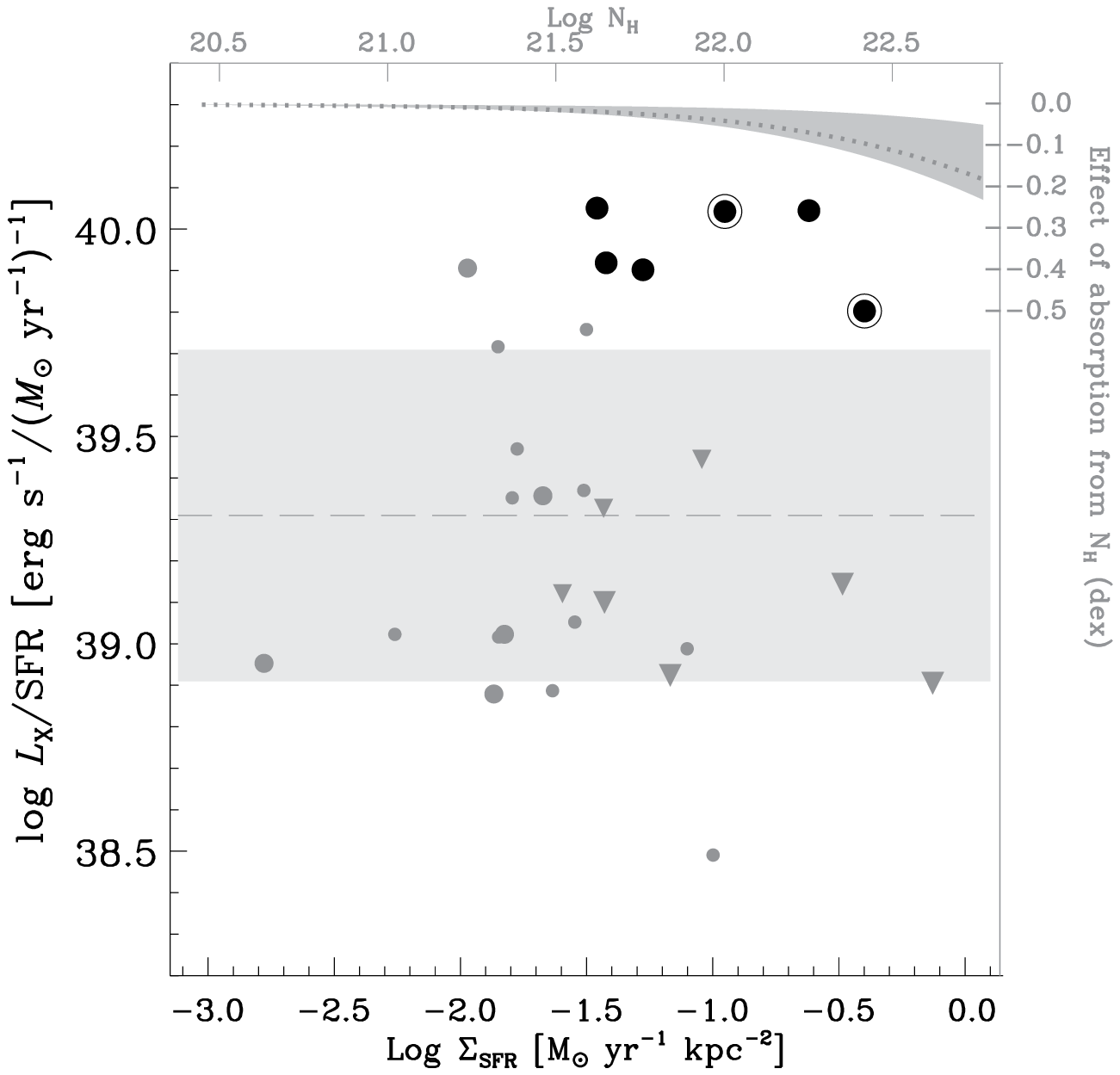}
     \end{center}
  \caption{Using IRX ({\em left panel}) and SFR surface densities ($\Sigma_{\rm SFR}$, {\em right panel}) to estimate the amount of neutral hydrogen (top axis), we find that X-ray absorption has a negligible effect on \lx/SFR (dark gray region, with dotted gray curve corresponding to solar metallicity; right axis shows the effect of $N_{\rm H}$ on \lx/SFR in dex units).  {\em Left}: While the estimated $N_{\rm H}$ based on IRX shows negligible effect on \lx/SFR, \lx/SFR does appear to correlate slightly (at 86\% level) with IRX (solid black line). However, this is driven by the ULIRGs (see discussion in section \ref{sec:NH}). {\em Right}: The SFR surface density (SFR/kpc$^{-2}$) relates to the gas density, using the Schmidt-Kennicutt Law. As expected from the estimated $N_{\rm H}$ (dark gray dotted line and region), \lx/SFR is not affected significantly by $\Sigma_{\rm SFR}$, consistent with the observed lack of correlation. }\label{fig:NH}
   \end{figure*}

We return to the question whether metallicity differences between the LBA sample and other HMXB-dominated galaxies might drive the elevated \lx/SFR. In Figure \ref{fig:metal} we show \lx/SFR vs. gas phase metallicity for the LBAs (black points) and the partial comparison sample (symbols are described in legend of Figure \ref{fig:lxsfr}). The data are correlated at the 99.2\% significance level,% [*** ALSO TRY: MULTIPLE CORRELATION TEST WITH NH***] 
based on a Spearman's rank correlation test.  Optical spectral data are available for the full LBA sample, but only 40\% of the HMXB-dominated (with sSFRs $>10^{-10}$ yr$^{-1}$) comparison sample.

Based on X-ray binary population synthesis models, the formation of HMXB populations is strongly metallicity-dependent \citep[\eg][]{Dray06, Linden2010, Fragos12, Fragos13}. One explanation is that stellar winds are weaker in massive stars with lower metallicities, allowing them to retain most of their mass and end their evolution as black hole companions. Since black hole HMXBs can be more luminous than neutron star HMXBs, the luminosity of a HMXB per SFR is expected to be higher in low-metallicity environments \citep{Mapelli09, ZamRob09, Fragos13}. In addition, a reduced stellar wind would affect the secular evolution of the binary and would result in less orbital expansion due to angular momentum loss from the stellar wind, and in overall more systems that will encounter Roche-lobe overflow mass transfer instead of Bondy-Hoyle type wind accretion. Roche-lobe overflow systems can drive much higher accretion rates onto the black hole and hence be more luminous X-ray sources \citep{Fragos12}. Indeed, initial results have shown evidence for an increased number of HMXBs, or off-nucleus ultraluminous X-ray sources (ULXs), per SFR in lower metallicity environments \citep{Swartz08, Mapelli09, Mapelli2010, Mapelli2011, Kaaret, Prestwich2013}. 

While the lower metallicities in the LBA sample are not as extreme as in dwarf galaxies, \citet{Overzier2010} find that $z\sim0.2$ LBAs have lower metallicities per stellar mass compared to the local mass-metallicity relation \citep[\eg][]{VAC_metal}, but consistent to the mass-metallicity relation of LBGs at $z\sim2$ \citep{erb06}. We estimate metallicities for the $z=1.9$ and 2.5 LBG samples from the X-ray stacking analysis in \citet{me-lbgstack} using the $z\sim2$ mass-metallicity relation (\citealt{erb06}; converting the metallicities to PP04 O3N2 units, using the conversions given in \citealt{KewleyEllison08}). The range of \lx/SFR vs. metallicity values for $z\sim2$ LBGs is shown as a blue hatched region in Figure \ref{fig:metal}. Based on the theoretical predictions that metallicity affects the number of luminous HMXBs, we postulate that the higher X-ray luminosity per SFR in the LBAs may be attributed to a higher number of more luminous HMXBs.

The gray line in Figure \ref{fig:metal} shows the theoretical prediction, based on X-ray binary population synthesis models \citep{Fragos13}. It is notable that these models predict that the X-ray luminosity from HMXBs per unit of SFR increases by approximately an order of magnitude going from solar metallicity to less than 10\% solar. While theoretical models express metallicity in absolute $Z$ units, we convert these to units of oxygen abundance ($12+\log$ [O/H]) by scaling solar metallicity ($Z_\odot$) to $12+ \log$ [O/H] $\sim$8.69. Our analysis shows reasonable agreement with the theoretical prediction (gray curve), albeit with significant scatter, which is likely attributed to the issues discussed below. 
 
We mention a couple of caveats to this analysis. First, we expect that the metallicities, measured using observed emission lines, are characteristic of star-forming regions where these lines are produced and HMXBs are presumed to be present. However, we are measuring the {\it galaxy-averaged} metallicity rather than the metallicities for the regions directly associated with HMXBs. The second caveat is that there is significant scatter between different techniques that measure metallicity and the absolute scaling is not well-constrained \citep[see \eg][ and discussion in Section \ref{sec:logoh}]{KewleyEllison08}, which complicates comparisons with theoretical models. Nevertheless, since the former issue adds scatter via stochastic variations between the galaxies and the latter introduces systematic errors, the observed correlation shows promising evidence that metallicity may be responsible for driving much of the scatter in the \lx/SFR relation, specifically offering a plausible explanation for the elevated \lx/SFR in the lower-metallicity LBAs. 

In Figure \ref{fig:LBA3}, the images of extended X-ray emission in three LBAs (VV~114, Haro~11 and J082355.0$+$280621.8) display multiple off-nuclear point sources with \lx$\gtrsim10^{40}$\ergs, presumed to be ULXs, associated with each galaxy. Based on their SFRs and the analysis of HMXBs in star-forming galaxies by \cite{Mineo12}, we would have expected fewer ULXs with \lx~$>10^{40}$ \ergs in these sources than what we observe (five, three and three off-nuclear sources with \lx$>10^{40}$ \ergs in VV~114, Haro~11and J082355.0$+$280621.8, respectively). Accounting for possible source confusion, where multiple sub-ULX sources appear as a single, unresolved \lx$>10^{40}$\ergs ULX, implies an improbably high number density of HMXBs per SFR. While further analysis is required to confirm this result, the numbers of ULXs suggest that LBAs host an excess of extremely luminous ULXs, potentially because they have uniquely high SFRs and low metallicities. 
   
\subsection{Effects of dust extinction}\label{sec:NH}
Typically, local galaxies appear to have higher levels of dust attenuation with increasing SFR \citep{Wang96, Martin05, ben}, but the LBA sample appears to be a unique population that is offset towards lower IRX values per SFR \citep{me-radio, Overzier2011}. We explore the possibility that LBAs appear to have elevated X-ray luminosities per SFR caused by dust effects. Since dust attenuation is uniquely low in LBAs, we might expect that typical galaxies suffer higher levels of dust obscuration, which might affect the observed \lx/SFR, \ie if X-ray emission were significantly absorbed in other star-forming galaxies, such as IR-selected samples with similar SFRs. 
In Figure \ref{fig:NH}, we show our results for testing the hypothesis that \lx/SFR depends on the amount of dust and gas, measured using two different methods: IRX (see Equation \ref{eqn:irx}) to measure the dust attenuation (shown in left panel), and the SFR surface density, $\Sigma_{\rm SFR}$, which is related to the gas surface density (\citealt{Kennicutt98}; henceforth referred to as the Schmidt-Kennicutt Law, shown in the right panel). 

In the left panel of Figure \ref{fig:NH}, we find there is a weak correlation between \lx/SFR and IRX, at a 86.4\% confidence level, if we restrict our analysis to the HMXB-dominated (with sSFR$>10^{-10}$yr$^{-1}$) galaxies with measured metallicities (the partial comparison sample, 38 galaxies, and six LBAs). The correlation is much stronger (99.0\%) if we include all of the HMXB-dominated galaxies (\ie including galaxies without measured metallicities, 76 galaxies from the comparison sample and six LBAs). Since dust and metallicity are also correlated in our data (at $\sim$95\% confidence level; see also \citealt{Skibba2011}), we consider the expected impact that dust extinction alone, using IRX as a proxy, might have on \lx/SFR. We estimate the neutral hydrogen column density from IRX using the following steps: 1. We convert IRX into the FUV attenuation ($A_{\rm FUV}$) following the relation given by Equation~15 in \citet[][]{Hao2011}: $A_{\rm FUV}=2.5 \log [1+0.46\times10^{\rm IRX}]$; 2. We estimate the attenuation in the $V$-band from $A_{\rm FUV}$ as $A_V\approx0.4 A_{\rm FUV}$, using the extinction laws from \cite{GdP}: $A_{\rm FUV}=7.9 E(B-V)$ and \cite{Cardelli89}:  $A_{V}=R_V E(B-V)$, where the V-band reddening is $R_{V}=3.1$; 3. According to \cite{GuverOzel}, the neutral hydrogen column density ($N_{\rm H}$) relates to the V-band attenuation as: $N_{\rm H}=2.21\times10^{21}  A_{V}$ cm$^{-2}$. Based on these steps, we derive a relation between $N_{\rm H}$ and IRX as:
\begin{equation}
N_{\rm H} \approx 2.21 \times 10^{21} [0.98 \log (1+0.46 \times 10^{\rm IRX})] {\rm cm}^{-2} \label{eqn:nh_irx}
\end{equation}
The top axis in Figure \ref{fig:NH} (left panel) displays the corresponding neutral hydrogen column density and the gray dotted curve marks the nearly negligible effect ($<0.05$ dex, see right axis) of this X-ray absorption on \lx/SFR. Since the X-ray absorption is a product of the metallicity and hydrogen column density (our dotted curve assumes solar metallicity), the same values of $N_{\rm H}$ have a stronger effect on the level of X-ray absorption with increasing metallicity. The dark gray region shows the range of X-ray absorption expected given the observed range of metallicities in our samples, 12$+\log$[O/H]$=8.1$(top edge)--8.85(bottom edge). 

Since IRX provides a global measure of the dust attenuation averaged over the full extent of the galaxy, by comparing the infrared emission to the ultraviolet emission, the IRX value may not accurately capture the true extinction present along the lines-of-sight to specific star-forming regions where HMXBs reside. Therefore, we try another independent measure to quantify the extinction in these galaxies. 

The SFR surface density (SFR/kpc$^{2}$, $\Sigma_{\rm SFR}=$SFR/$\pi ab$, where $a$ and $b$ are the semi-major and semi-minor axes), is related to neutral gas surface density ($\Sigma_{\rm H}$) as $\Sigma_{SFR}=2.5\times10^{-4} [\frac{\Sigma_{\rm H}}{M_{\odot} \rm pc^{-2}}]^{1.4}$ \msun yr$^{-1}$ kpc$^{-2}$ \citep{Kennicutt98}. The neutral hydrogen gas column density is $\Sigma_{\rm H}/m_{\rm H}$, where $m_{\rm H}$ is the mass of a hydrogen atom. Therefore, we estimate $N_{\rm H}$ from the UV$+$IR-derived SFR surface density by:
\begin{equation}
N_{\rm H} = 4.6\times10^{22} ~\Sigma_{\rm SFR}^{0.7} {\rm cm}^{-2} \label{eqn:nh_sigmasfr}
\end{equation}

We see no correlation of \lx/SFR with $\Sigma_{\rm SFR}$ (see right panel of Figure \ref{fig:NH}), as expected based on the small effect of $N_{\rm H}$ (derived from $\Sigma_{\rm SFR}$, as given in Equation \ref{eqn:nh_sigmasfr}; gray dotted curve) on \lx/SFR. However, the Schmidt-Kennicutt Law may not apply to galaxies with high SFR efficiencies, \ie galaxies with high SFR surface density per gas density \citep[\eg LBAs; ][ and possibly LIRGS; \citealt{Garcia2012}]{me-thesis}, which explains why the estimated $N_{\rm H}$ values from the two techniques differ. However, the two techniques do provide comparable values of $N_{\rm H}$ for the other star-forming galaxies (gray points in Figure \ref{fig:NH}). 

Neither measure of extinction implies sufficient $N_{\rm H}$ to significantly absorb 2--10~keV X-ray emission. However, since both IRX and $\Sigma_{\rm SFR}$ are calculated using global properties of the galaxies, we are measuring the average absorption and are unable to account for heavily obscured sub-galactic regions. For example, in ULIRGs, \citet{GenzelLIRG} find incredibly high optical absorption values, $A_{V}\approx 50-1000$~mag for some ULIRGs, where the optical dust column densities are comparable to the dense molecular cloud (CO) millimeter observations. Therefore, HMXBs found in such heavily obscured regions would be affected by significant X-ray absorption and in such specific cases the observed X-ray luminosities may be underestimated. While this scenario is plausible in some ULIRGs and LIRGs, X-ray absorption is not likely to fully explain the apparent correlation between \lx/SFR and IRX across the entire population of star-forming galaxies. However, the correlation between \lx/SFR and IRX is mainly influenced by the ULIRGs; \ie the data are not correlated ($<50\%$ significance level) when ULIRGs are eliminated.  

%While we have assumed that the ratio of IR to UV (IRX) measures dust attenuation, we mention the additional complication that the IR- and UV-emission probe slightly different timescales. Specifically, $\sim30$ Myr after a burst of star-formation, the dusty birth cloud enshrouding the star-forming region starts to dissipate, allowing UV photons to escape more freely \citep{Calzetti07}. Since only a small fraction (the most massive) of HMXBs form within 30 Myr, X-ray emission may relate better to the timescales corresponding to when UV emission can escape \citep[ $>30$ Myr; \eg see][]{Mineo13}.

\section{Implications for X-ray emission in high redshift galaxies}\label{sec:highz}
While in this study we are studying the X-ray emission in low redshift ($z<0.1$) analogs of LBGs, a parallel aim is to gain some physical insight about the important factors that affect X-ray emission in distant LBGs. Since individual X-ray detections of LBGs at $z>1$ are likely dominated by AGN, the study of XRB populations in these galaxies is limited to their average X-ray emission based on stacking of large numbers of galaxies. \cite{me-lbgstack} perform stacking of $z=1.5$--5 LBGs in the 4~Ms CDFS and find evolution in the \lx/SFR ratio over the history of the universe as described by the following equation: $\log L_{\rm X}=0.90\log (1+ z) + 0.65\log {\rm SFR} + 39.8$. \citet{Fragos12}, combining X-ray binary population synthesis models with results from the Millennium II simulation and semi-analytical galaxy catalogs \citep{Guo}, predict similarly that the X-ray emission from HMXBs per unit SFR evolves with redshift, driven by the redshift evolution in the mean stellar metallicity. In this paper, we arrive at the same conclusion using individual LBAs.

Based on the numbers of resolved HMXBs within a few extended LBAs, we find additional clues that these lower-metallicity, high SFR galaxies contain a higher frequency of ULXs with \lx$>10^{40}$\ergss. Investigating the luminosity distributions of HMXBs in LBGs is impossible, due to the high spatial resolution and sensitivity requirements. However, extrapolating our result to higher redshift LBGs, we might expect that galaxies in the early Universe hosted a larger number of ULXs than predicted based on typical star-forming galaxies \citep[\eg][]{Mineo12}.

\section{Summary and Future Work}\label{sec:conclusion}
In this paper, we present our results on the X-ray emission from six LBAs in the local Universe ($z<0.1$). Whereas our knowledge about the high energy emission from star-forming galaxies in the early universe ($z\gtrsim1$) is limited by statistical techniques such as stacking, in this paper, we study the X-ray emission in {\it individual} galaxies from a sample of $z<0.1$ galaxies that are excellent analogs to distant LBGs. Based on our comparison of LBAs with other star-forming galaxies, we summarize our main results as follows: 
%Whereas the study of X-ray emission from star-forming galaxies is restricted to stacking analysis at $z\gtrsim1$, these low redshift ($z<0.3$) analogs offer insight about the X-ray emission from individual galaxies and require much shorter exposure times to detect. 
\begin{itemize}
\item{We find that the hard (2--10~keV) X-ray emission per unit SFR is elevated by $\sim 1.5\sigma$ in the LBA population, similar to X-ray stacking results from $z>3$ LBGs (Figure \ref{fig:lxsfr}). }
\item{We find that the lower metallicities in the LBA sample may be driving the elevated \lx/SFR  (Figure \ref{fig:metal}). }
\item{We also find that this effect is {\it not} driven by extinction as we find no significant correlation of \lx/SFR with IRX or SFR surface density and the inferred column densities are too low to cause significant absorption in the 2--10~keV X-ray luminosity (Figure \ref{fig:NH}).}
\end{itemize}

Our study of X-ray emission in LBAs, a unique population of galaxies having high SFR and low metallicities, highlights the importance of metallicity on the scaling between HMXB X-ray emission and SFR. However, this result is limited by the small sample size; a sample spanning a wider range of SFRs, metallicities, and dust attenuation values would better constrain the physical properties responsible for driving the elevated \lx/SFR observed in the current LBA sample. 

We find tentative evidence that LBAs, possibly owing to their high SFRs and low metallicities, host a significant excess of $>10^{40}$ \ergs ULXs. This result is based on \Chandra~observations of individually resolved X-ray sources within two extended LBAs, VV114 and J082355.0$+$280621.8. However, the data for J082355.0$+$280621.8 were not complete at the $10^{40}$ \ergs limit and require deeper follow-up observations. Additional observations of spatially extended LBAs, only possible with the high angular resolution capabilities of \Chandra, would provide better statistics on the distribution of XRB populations, in order to study the high luminosity tail of the X-ray luminosity function (XLF) which can be compared to XLFs of local star-forming galaxies \citep[\eg][]{Mineo12} and X-ray population synthesis models \citep[see, \eg][]{Tzanavaris12}.

While the ability to study X-ray binary populations in individual high redshift ($z>2$) LBGs may come to fruition in the future (\eg an X-ray telescope with large collecting area such as {\em SMART}-X), these $z<0.1$ analogs offer some important clues suggesting that the relationship between \lx~and SFR evolves with the chemical evolution of the Universe. In this paper, we establish that such a relationship is not merely occurring in an average sense over cosmic time \citep{me-lbgstack, Fragos12}, but that the scatter in the \lx/SFR relation may be attributed to metallicity variations between individual galaxies. 

\begin{acknowledgments}
This research was supported by \Chandra~Cycle 12 program \#12620841 (P.I.: Basu-Zych). We thank A. Prestwich for sharing her work on ULXs in extremely low metallicity galaxies, S. Mineo for providing useful comparison data, and J. Mullaney for sharing his IR code. The author gratefully acknowledges B.D. Johnson, A. Henry, and V. Antoniou for helpful discussions. T. F. acknowledges support from CfA and ITC prize fellowship programs.  This publication makes use of data products from the Wide-field Infrared Survey Explorer, which is a joint project of the University of California, Los Angeles, and the Jet Propulsion Laboratory/California Institute of Technology, funded by the National Aeronautics and Space Administration.
\end{acknowledgments}

\bibliographystyle{/Users/Antara/research/latex/apj}
\bibliography{/Users/Antara/research/latex/apj-jour,/Users/Antara/research/latex/antara_refs,/Users/Antara/research/latex/library_mod,/Users/Antara/research/latex/xray_lba,/Users/Antara/research/UVLG2010/prop_chandra/ms,/Users/Antara/research/latex/xray_stacking}

\end{document}